\documentclass[iop]{emulateapj}

\usepackage{amsmath,amssymb}
\usepackage{color}
\usepackage{graphicx}
\usepackage{natbib}
\usepackage{hyperref,soul,nicefrac}

\newcommand{\ra}{\textit{RadioAstron}}

\makeindex
\citeindextrue

\received{January 21, 2016}
\accepted{March 1, 2016}
\journalinfo{Astrophysical~journal Letters}

\shorttitle{Extreme brightness temperature of 3C\,273}
\shortauthors{Kovalev et al.}

\begin{document}
\title{\textit{RadioAstron} Observations of the Quasar 3C\,273: a Challenge to the Brightness Temperature Limit}
\author{
Y. Y. Kovalev\altaffilmark{1,2},
N. S. Kardashev\altaffilmark{1},
K. I. Kellermann\altaffilmark{3}, 
A. P. Lobanov\altaffilmark{2,4}, 
M. D. Johnson\altaffilmark{5}, 
L. I. Gurvits\altaffilmark{6,7}, 
\mbox{P. A. Voitsik\altaffilmark{1}},
J. A. Zensus\altaffilmark{2}, 
J. M. Anderson\altaffilmark{2,8}, 
U. Bach\altaffilmark{2},
D. L. Jauncey\altaffilmark{9,10}, 
F. Ghigo\altaffilmark{11}, 
T. Ghosh\altaffilmark{12},
A. Kraus\altaffilmark{2}, 
Yu. A. Kovalev\altaffilmark{1}, 
M. M. Lisakov\altaffilmark{1},
L. Yu. Petrov\altaffilmark{13}, 
J. D. Romney\altaffilmark{14}, 
C. J. Salter\altaffilmark{12},
K. V. Sokolovsky\altaffilmark{1,15}
}

\altaffiltext{1}{Astro Space Center of Lebedev  Physical Institute, Profsoyuznaya 84/32, 117997 Moscow, Russia
}
\altaffiltext{2}{Max-Planck-Institute for Radio Astronomy, Auf dem H\"ugel 69, D-53121, Germany}
\altaffiltext{3}{National Radio Astronomy Observatory, 520 Edgemont Road, Charlottesville, VA 22903-2475, USA}
\altaffiltext{4}{Institut f\"ur Experimentalphysik, Universit\"at Hamburg, Luruper Chaussee 147, 22761 Hamburg, Germany}
\altaffiltext{5}{Harvard-Smithsonian Center for Astrophysics, 60 Garden Street, Cambridge, MA 02138, USA}
\altaffiltext{6}{Joint Institute for VLBI ERIC, PO Box 2, 7990 AA Dwingeloo, The Netherlands}
\altaffiltext{7}{Department of Astrodynamics and Space Missions, Delft University of Technology, 2629 HS Delft, The Netherlands}
\altaffiltext{8}{Helmholtz-Zentrum Potsdam, Deutsches GeoForschungsZentrum, Department 1: Geodesy and Remote Sensing, Telegrafenberg, 14473, Potsdam, Germany}
\altaffiltext{9}{CSIRO Astronomy and Space Sciences, Epping, NSW 1710, Australia, Australia}
\altaffiltext{10}{Research School of Astronomy and Astrophysics, Australian National University, Canberra, ACT, 2611, Australia}
\altaffiltext{11}{National Radio Astronomy Observatory, Rt. 28/92, Green Bank, WV 24944-0002, USA}
\altaffiltext{12}{Arecibo Observatory, NAIC, HC3 Box 53995, Arecibo, Puerto Rico, PR 00612, USA}
\altaffiltext{13}{Astrogeo Center, 7312 Sportsman Dr., Falls Church, VA 22043, USA}
\altaffiltext{14}{National Radio Astronomy Observatory,  P.O. Box O, 1003 Lopezville Road, Socorro, NM 87801-0387, USA}
\altaffiltext{15}{Sternberg Astronomical Institute, Moscow State University, Universitetskii prospekt 13, 119992 Moscow, Russia}

\begin{abstract}
Inverse Compton cooling limits the brightness temperature of the radiating plasma to a maximum of $10^{11.5}$~K. Relativistic boosting can increase its observed value, but apparent brightness temperatures much in excess of $10^{13}$~K are inaccessible using ground-based very long baseline interferometry (VLBI) at any wavelength. We present observations of the quasar 3C\,273, made with the space VLBI mission \ra\ on baselines up to 171,000~km, which directly reveal the presence of angular structure as small as 26~$\mu$as (2.7 light months) and brightness temperature in excess of $10^{13}$~K. These measurements challenge our understanding of the non-thermal continuum emission in the vicinity of supermassive black holes and require a much higher Doppler factor than what is determined from jet apparent kinematics.
\end{abstract}

\keywords{
galaxies: active --- 
galaxies: jets --- 
radio continuum: galaxies --- 
techniques: interferometric --- 
quasars: individual (3C\,273)
}

\section{Introduction}
\label{s:intro}

The discovery, half a century ago, of quasar radio flux density variability on timescales as short as months \citep{shol65_2,shol65_1,dent65} indicated linear sizes as small as light months corresponding to angular dimensions of the order of one mas or less and suggesting extraordinarily high apparent brightness temperatures \citep{PTK66}. The subsequent discovery with VLBI of superluminal motion and relativistic beaming with Doppler factors $\delta\sim10$ \citep[e.g.,][]{lister_etal13} appeared to explain the extreme values of apparent size and brightness temperature inferred from intrinsic variability \citep{KPT81}. For incoherent synchrotron emission of relativistic electrons, in the absence of relativistic boosting, the maximum brightness temperature is limited by inverse Compton cooling to about $10^{11.5}$~K \citep{KPT69,R94} or $5\times10^{10}$~K if there is equilibrium between the energy in particles and magnetic fields \citep{R94}. Relativistic boosting can increase the observed brightness temperature, $T_\mathrm{obs}=\delta \times T_\mathrm{int}$ by the Doppler factor $\delta=(1-\beta^2)^{\nicefrac{1}{2}}(1-\beta\cos\varphi)^{-1}$, where $T_\mathrm{int}$ is the intrinsic brightness temperature, $\beta=v/c$ is the jet speed in units of the speed of light, and $\varphi$ is the angle between the jet velocity vector and the line of sight. The apparent jet speed $\beta_\mathrm{app}=\beta\sin\varphi(1-\beta\cos\varphi)^{-1}$.

However, observations of interstellar scintillation in many flat-spectrum sources \citep[e.g.,][]{lovell_etal08}, reflecting very small angular sizes and very high brightness temperatures, as well as rapidly varying TeV emission from some quasars \citep{HH09}, suggested either much higher Doppler factors than those estimated from VLBI observations of quasar kinematics \citep{lister_etal13}, or possibly a different emission mechanism than incoherent synchrotron radiation from relativistic electrons \citep[e.g.,][]{BlandfordKonigl79}. Previous ground-based \citep{2cmPaperIV,Lee_etal08} and space-based \citep{fre02,Tingay_etal01,VSOPsurvey5} VLBI (SVLBI) observations are consistent with the expected small sizes and high observed brightness temperatures of quasars deduced from their intrinsic variability as well as with the limits discussed above. However, previous VLBI measurements could not attain the baseline length or angular resolution needed to address the question of high brightness temperature conclusively.

\begin{figure*}[t]
\begin{center}
\includegraphics[width=0.39\textwidth,trim = 0cm 0cm 0cm 0cm]{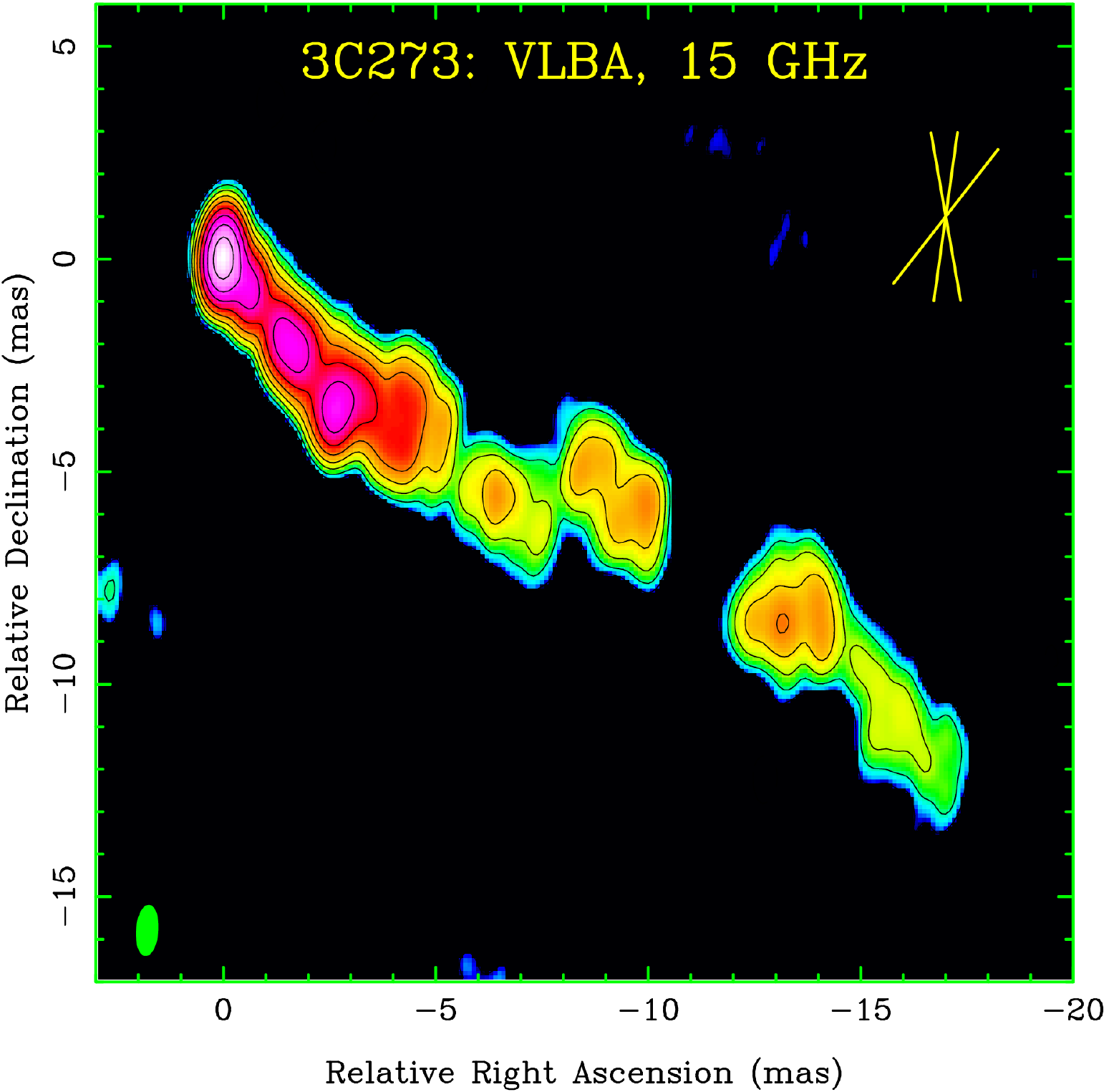}
\includegraphics[width=0.56\textwidth,trim = -0.1cm 0cm 0cm 0cm]{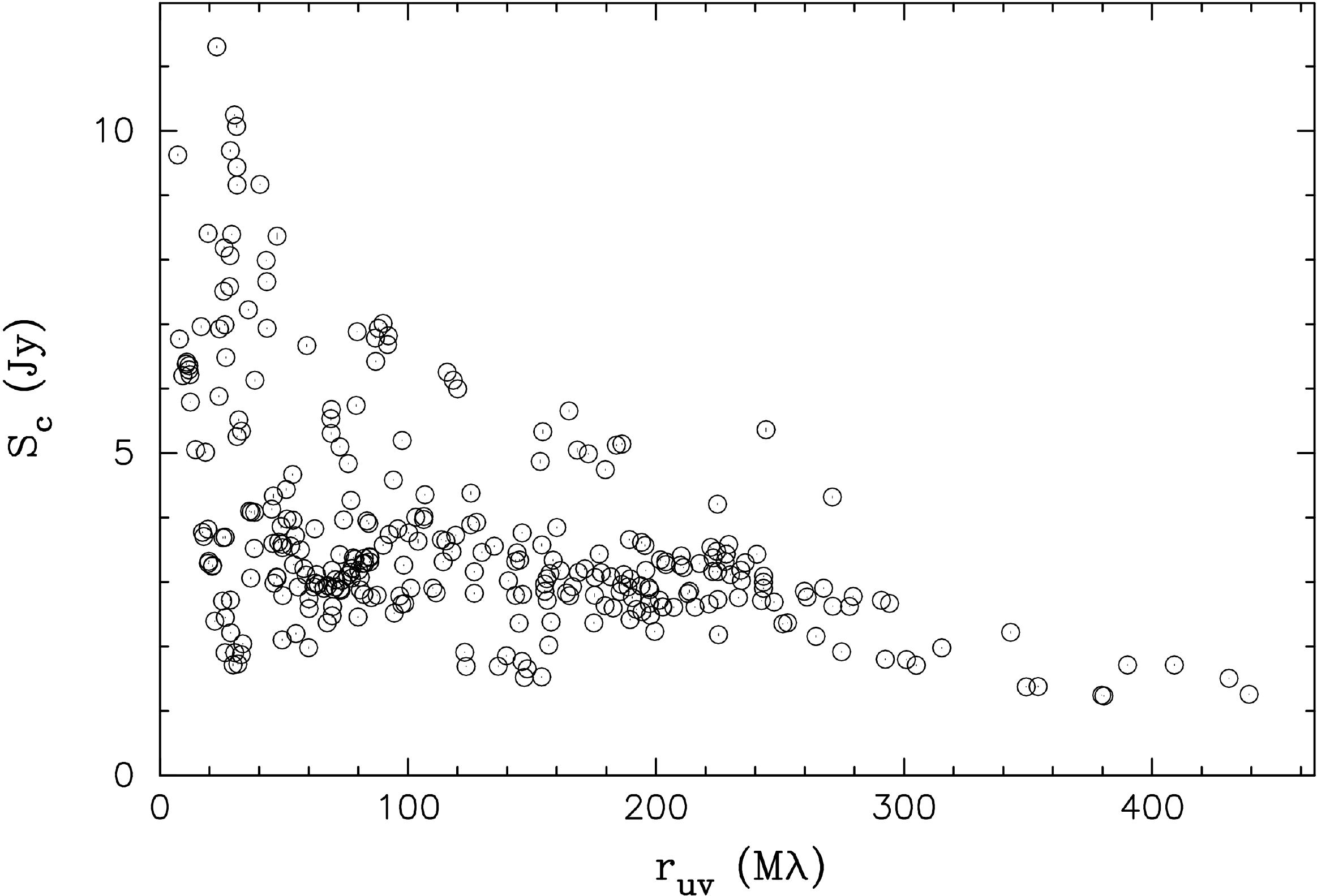}
\includegraphics[width=0.39\textwidth,trim = 0cm 0cm 0cm -1cm]{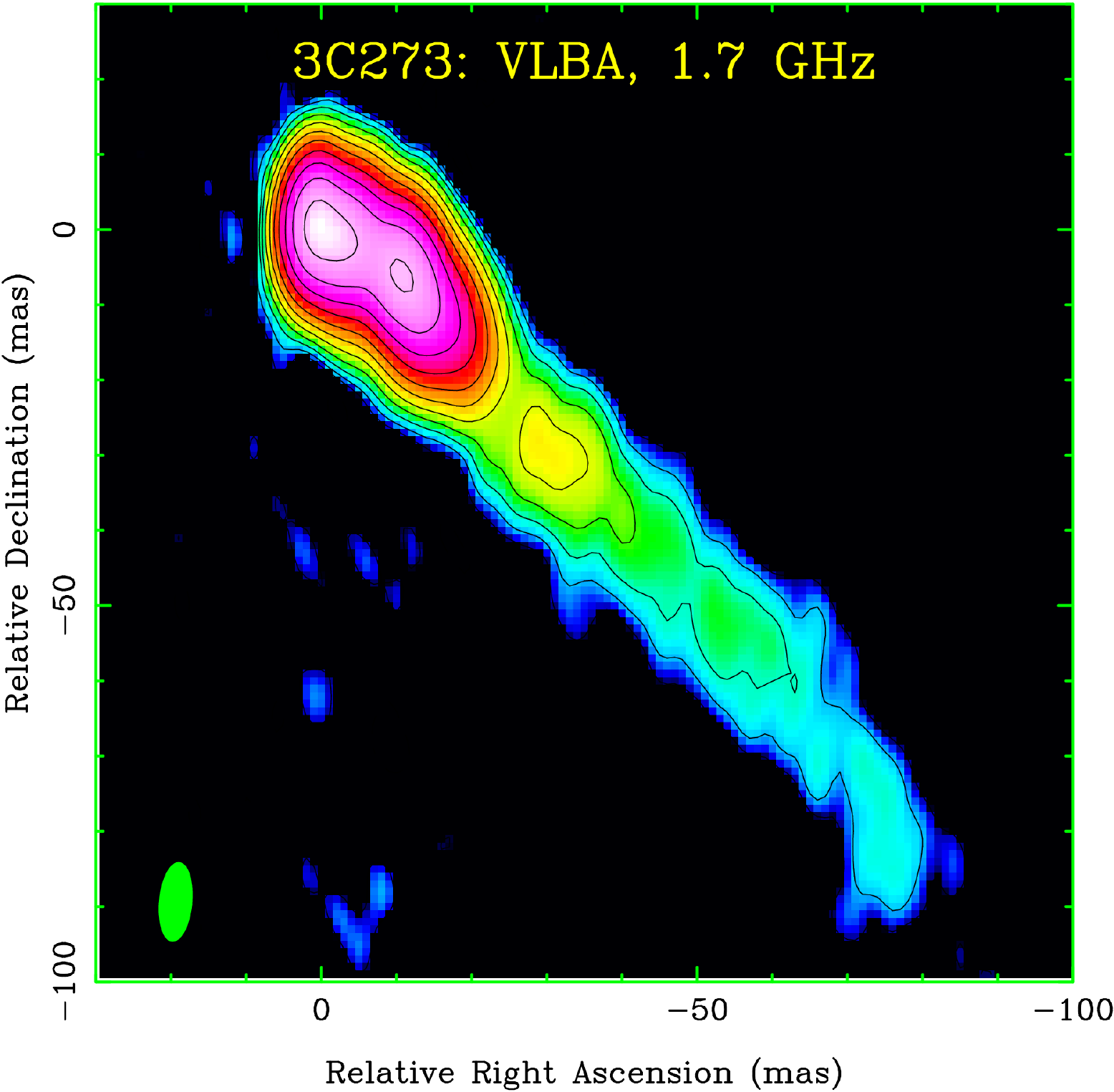}
\includegraphics[width=0.56\textwidth,trim = -0.1cm 0cm 0cm -1cm]{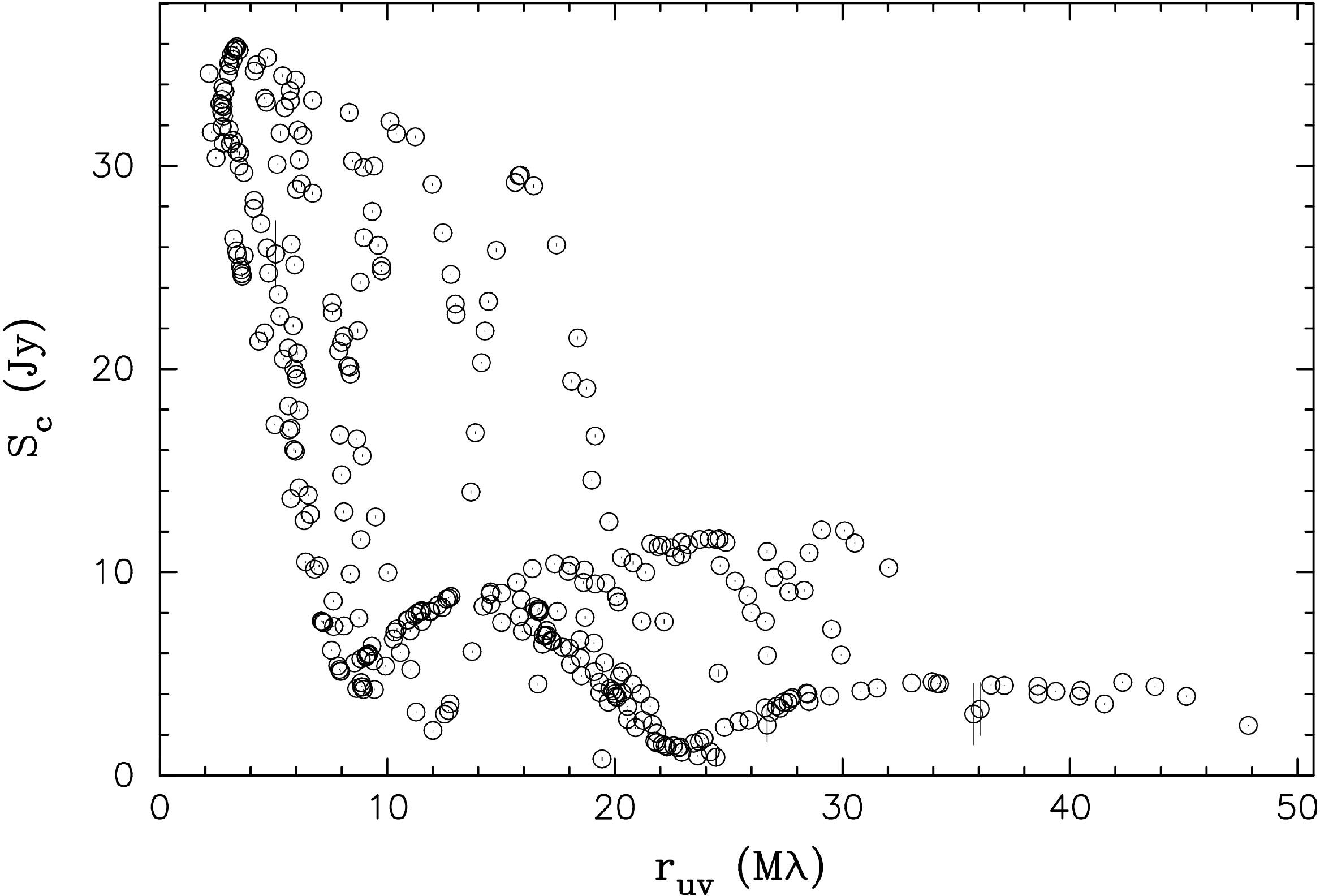}
\end{center}
\caption{\label{f:ground}
Results of ground-based VLBA observations of the quasar 3C\,273 made at 15~GHz (top, 10~February 2013, peak intensity 3.2~Jy/beam, noise level 1.4~mJy/beam, beam size $1.21\times0.53$~mas) and 1.6 GHz (bottom, 18~August~2011, peak intensity 9.6~Jy/beam, noise level 2.2~mJy/beam, beam size $10.6\times4.4$~mas). 
\textit{Left}: Total intensity pseudo-color and contour images with the beam shown at the half power level in green. Contours start at 0.25\,\% and 0.07\,\% of the peak intensity and are plotted with $\times2$ steps at 15 and 1.6~GHz, respectively.
The yellow lines in the 15~GHz image show the range of position angles of ground-to-space \ra\ baselines covered by our observations: $10^\circ$, $-8^\circ$, $-38^\circ$.
\textit{Right}: Corresponding interferometric (fringe) amplitude versus projected baseline.
}
\end{figure*}

With the launch by the Russian Federal Space Agency of the \ra\ Space Radio Telescope (SRT) on July 18, 2011, it is now possible to test directly the regime of extremely high apparent brightness temperatures. The ``Spektr-R'' satellite is in a nine-day highly elliptical orbit around the Earth with a variable apogee up to 350,000~km. It carries a 10-meter parabolic radio telescope and operates at 330~MHz (92~cm), 1.7~GHz (18~cm), 4.8~GHz (6.2~cm), and 22~GHz (1.3~cm) \citep{ra1}.

The quasar 3C\,273 has a redshift of $z=0.158$, is at a luminosity distance of 750~Mpc, and has been regularly imaged over the past four decades, including observations at 15~GHz with the Very Long Baseline Array (VLBA) as part of the MOJAVE survey \citep{lister_etal13}. As seen in Fig.~\ref{f:ground}, the jet rapidly becomes well resolved. However, the fringe amplitude remains near several Jy out to maximum ground baselines. This corresponds to the bright feature at the apparent base of the jet where the optical depth approaches unity. Previous brightness temperature estimates for the core of 3C\,273 include $5\times10^{11}$~K from space-based \citep{VSOPsurvey5} and up to $6\times10^{12}$~K from ground-based \citep{2cmPaperIV} VLBI experiments.

Here we report the results of space VLBI observations of 3C\,273 at 1.3, 6.2, and 18 cm wavelengths on interferometer baselines from \ra\ to the NRAO 100-meter Robert C.\ Byrd Green Bank Telescope (GBT), the phased Karl G.\ Jansky Very Large Array (VLA), the 100-meter Effelsberg radio telescope, and the 305-meter Arecibo Observatory William E.\ Gordon Radio Telescope in the December~2012 --- February~2013 period. The projected interferometer baselines ranged up to 171,000~km and $7.6\times10^9\lambda$ (see Table~\ref{t:results} for details).
We note that later \ra\ observations of 3C\,273 at 18, 6.2, and 1.3~cm up until the middle of 2015 have not delivered any detections for projected baselines longer than 1.1~G$\lambda$.
We adopt a cosmology with $\Omega_m = 0.27$, $\Omega_\Lambda = 0.73$ and $H_o = 71 \; \mathrm{km\; s^{-1} \; Mpc^{-1}}$ \citep{WMAP5_COSMOLOGY}.

\section{\ra\ Space VLBI observations, correlation and signal search}

Observations were made either in two 16~MHz wide IF channels for the orthogonal circular polarizations in a given frequency band, or simultaneously in two frequency bands with a single polarization in each band. The central frequencies of observations were: 1668.0~MHz (L-band), 4836.0~MHz (C-band), and 22236.0~MHz (K-band). For the SRT, after conversion to baseband, the four 16~MHz IF bands were sampled at the Nyquist rate with 1~bit sampling. Samplers and local oscillators were controlled by highly stable hydrogen clocks located at every antenna, including the SRT. The 128~Mbps data stream was transmitted to Earth over a 15~GHz downlink to the 22-meter antenna located at the Pushchino Radio Astronomy Observatory near Moscow. After decoding, the data were stored and then sent over the network link to the Astro Space Center (ASC) in Moscow where they were cross-correlated with the data recorded with 2~bit sampling at Arecibo, Effelsberg, the GBT or the VLA using both ASC-developed \citep{ra1} and DiFX \citep{del11} software correlators.

In a space VLBI system, where the accuracy of the orbit determination is limited and the uncertainty of the baseline vector and its time derivatives is large, the interferometric response needs to be searched over a wide range of trial delays, delay rates, and delay accelerations. Post-correlation data analysis was performed using the PIMA software package \citep{VGaPS} including fringe fitting with compensation for all three terms. Fig.~\ref{f:K_fringe} shows a plot of fringe amplitude as a function of phase delay and fringe frequency for the SRT--GBT observations of 3C\,273 at 22~GHz on 2~February~2013. After applying the initial residual delay of 17.7~$\mu$s, fringe rate offset of 2.45~Hz, and residual acceleration of less than $3\times10^{-16}$~s/s$^{-2}$, the data were re-correlated so that the interferometric response appears close to the origin in Fig.~\ref{f:K_fringe}.

The interferometer response shown in Fig.~\ref{f:K_fringe}, as well as the SNR values presented in Table~\ref{t:results}, were integrated over 9.5~min. The probability of false detection for every measurement shown is estimated to be less than $10^{-12}$. Extensive checks have shown that none of the maser based local oscillator systems, either on the ground or on the spacecraft, limit the coherence over the 9.5 minute scan intervals. Interferometers with ground--ground baselines have coherence losses associated with random phase variations along the path through the atmosphere to both stations. Space-to-ground interferometers have coherence losses associated with propagation though the atmosphere to only one station, so the resulting coherence time using baselines to the SRT is longer than is typically achieved with ground-ground systems. Our analysis of many \ra\ observations has shown that after applying a residual orbit acceleration term to the data, there are no significant coherence losses within a 9.5~min integration at 6 and 18 cm in most experiments. A partial coherence loss occurs at 1.3~cm; it affects the fringe amplitudes and, consequenly, SNR measured by the baseline-based fringe fitting (Table~\ref{t:results}, column~6). The residual ground-based phase variations are risky to compensate in cases of degenerate ground-to-space telescope triangles with low SNR SVLBI baselines due to possible overestimation of the \ra\ fringe amplitude \citep[see for details][]{MVM08}. Because of that, we have used 200-s integrations to estimate correlated flux densities shown in column (8) of Table~\ref{t:results} since coherence analysis of ground-based data has shown no significant losses over this time interval.

%

\section{Brightness temperature}
\subsection{Brightness temperature estimates from single-component Gaussian models}

\begin{figure}[t]
\begin{center}
\includegraphics[width=0.47\textwidth,trim = 0cm 0cm 0cm 0cm,clip]{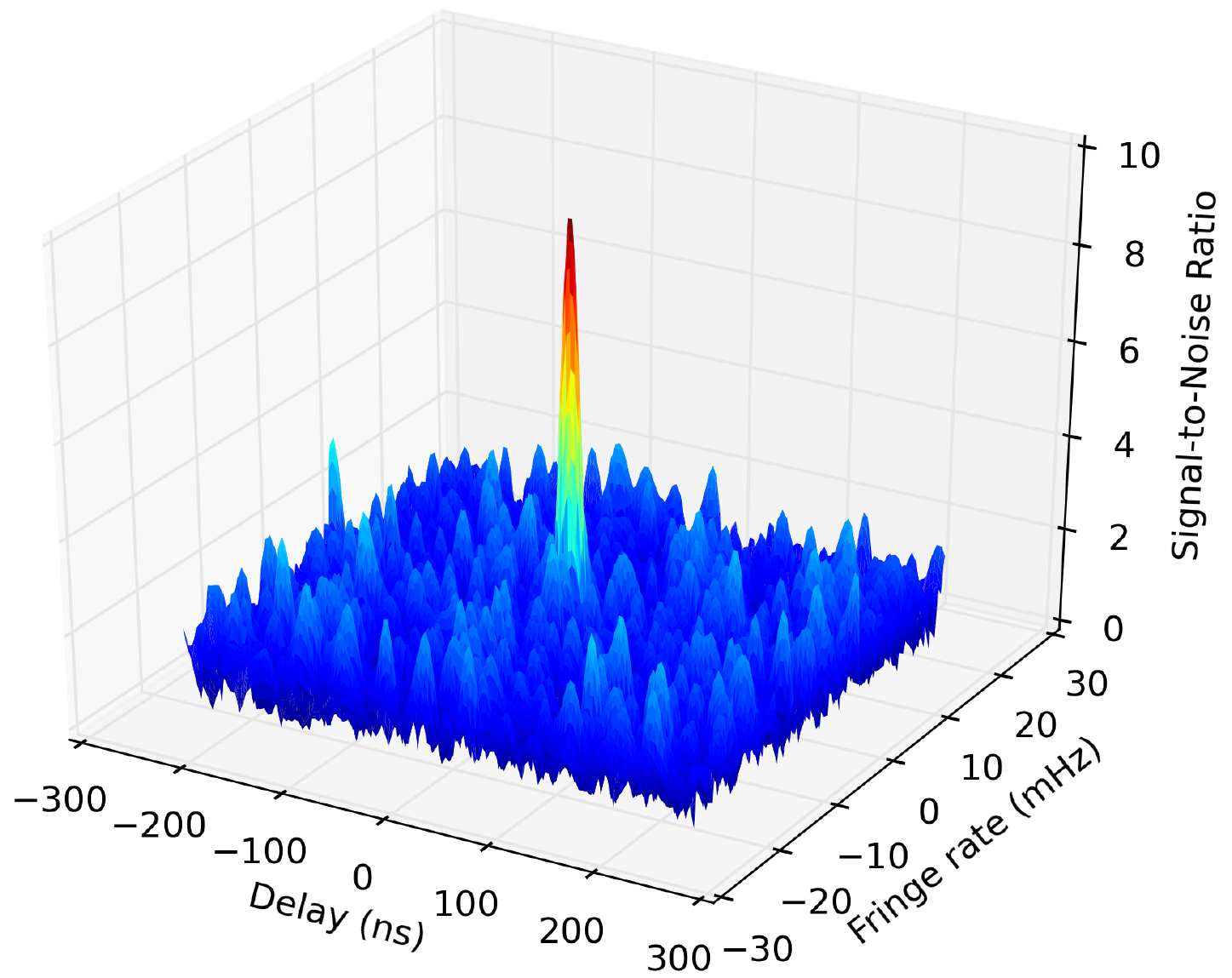}
\end{center}
\caption{\label{f:K_fringe}
\ra\ (Space Radio Telescope--GBT) 22~GHz interferometer response from observations of the quasar 3C\,273 on 2~February~2013 with a projected baseline of 7.6 G$\lambda$. Shown is the ratio of interferometer fringe amplitude to the average noise amplitude as a function of the residual delay (in nanoseconds) and fringe rate (in millihertz).
}
\end{figure}

\begin{deluxetable*}{cllrrrrrrcrr}
\tablecolumns{12}
\tabletypesize{\scriptsize}
\tablewidth{0pt}
\tablecaption{\label{t:results}\ra\ ground-to-space radio interferometer measurements of the quasar 3C\,273}
\tablehead{
\colhead{$\lambda$} & \colhead{Epoch} & \colhead{GRT} & \colhead{$r_{uv}$} & \colhead{P.A.} & \colhead{SNR} & \colhead{$S_\mathrm{t}$} & \colhead{$S_\mathrm{c}$} & \colhead{$\theta$} & \colhead{$T_\mathrm{b}$}& \colhead{$T_\mathrm{b,min}$} & \colhead{$T_\mathrm{b,char}$} \\
\colhead{(cm)} &&& \colhead{($10^3$\,km; G$\lambda$)} & \colhead{(deg)} && \colhead{(Jy)} & \colhead{(mJy)} & \colhead{($\mu$as)}& \colhead{($10^{12}$ K)} & \colhead{($10^{12}$ K)} & \colhead{($10^{12}$ K)} \\
\colhead{(1)} & \colhead{(2)} &\colhead{(3)} & \colhead{(4)} & \colhead{(5)}  & \colhead{(6)} & \colhead{(7)} & \colhead{(8)} & \colhead{(9)} & \colhead{(10)} & \colhead{(11)} & \colhead{(12)}
}
\startdata
1.3  & 2013-02-02 & Gb, Y27 & 103; 7.6~ &  $-7$ &  9.8 & 3.4 & $125\pm22$ &  26 &  14 & 5.3 & 12 \\
6.2  & 2012-12-30 & Ar, Ef  &  90; 1.45 &   10  & 18.6 & 4.3 & $125\pm17$ & 142 &  13 & 4.5 & 18 \\
6.2  & 2013-02-02 & Ar      & 103; 1.69 &  $-8$ & 11.6 & 4.3 & $123\pm19$ & 122 &  17 & 5.2 & 15 \\
18   & 2013-01-08 & Gb      & 157; 0.87 & $-32$ &  8.9 & 5.0 & $ 42\pm7$  & 275 &  34 & 4.0 & 10 \\
18   & 2013-01-25 & Ar, Gb  & 171; 0.95 & $-38$ & 12.0 & 5.0 & $ 52\pm9$  & 246 &  42 & 6.3 & 18
\enddata
\tablecomments{
(1) the wavelength of observation; 
(2) the date of observation shown as year-month-day; 
(3) the ground radio telescopes with which significant SVLBI fringes were detected (Ef: Effelsberg, Gb: the GBT, Y27: phased VLA, Ar: Arecibo); 
(4) the average projected interferometer baseline in the plane of the sky and in units of wavelength;
(5) the position angle of the projected baseline; 
(6) the highest signal-to-noise ratio of fringe detection observed in a 9.5~min integration; 
(7) estimated (see text for details) flux density of the core component;
(8) the highest measured interferometric (fringe) amplitude with its rms error for the given observing segment, the error estimate includes the amplitude calibration uncertainty following \cite{RAampcal2014}; 
(9) the estimated angular size of the observed structure, full width at half maximum for the assumed circular Gaussian brightness distribution;
(10) brightness temperature in the source frame corresponding to the size given in column (8);
(11 and 12) minimum and characteristic brightness temperature in the source frame obtained following the approach suggested by \cite{l15}.
}
\end{deluxetable*}

The observational data and estimates of core parameters are summarized in Table~\ref{t:results}. 
Both the February~2013 MOJAVE observations at 15~GHz (Fig.~\ref{f:ground}, taking the core flux density and assuming its radio spectrum to be flat) and our VLA--GBT results at 22~GHz provide the total flux density (``zero'' spacing) estimate of the compact core feature as $S_\mathrm{t}=3.4$~Jy. 
The 18~cm core flux density $S_\mathrm{t}=5.0$~Jy is estimated from modeling the 18~cm VLBA imaging data (Fig.~\ref{f:ground}) and assuming that its flux density did not change significantly over the 1.5 years since the VLBA observations. This is supported by comparing visibility data with our measurements within \ra\ experiments at similar ground baselines. The 6.2~cm core flux density $S_\mathrm{t}=4.3$~Jy is estimated by an interpolation of 18 and 1.3~cm values with a spectral index $\alpha\approx-0.15$ ($S\propto\nu^\alpha$).

At 1.3~cm we have two ground--space interferometer measurements, both at projected baselines of about 8~G$\lambda$ corresponding to an interferometer fringe spacing of 27~$\mu$as. Fringes were detected on \ra\ baselines to both the GBT and the VLA with signal-to-noise ratios (SNR) of about ten (see Fig.~\ref{f:K_fringe}). Measured fringe amplitudes on these baselines are close to 0.1~Jy.
Assuming a circular Gaussian model, this corresponds to dimensions of this 1.3~cm feature of about 26~$\mu$as or 2.7 light months. We emphasize that size and brightness temperature estimates presented in Table~\ref{t:results} are relatively insensitive to uncertainties in the zero spacing flux density and to errors in the measured fringe amplitude.

The longest projected physical baseline showing interferometric fringes was 171,000~km, from the \ra\ Space Radio Telescope to the GBT and Arecibo at 18~cm. The ability of an interferometer to measure the brightness temperature of a resolved source depends on the physical length of the interferometer baseline \citep[see for details][]{2cmPaperIV} but is independent of wavelength. Therefore, our 18~cm \ra\ observations probe the highest values of the brightness temperature. They show a fringe amplitude of about 50~mJy. The corresponding brightness temperature is $4\times10^{13}$~K. 
The 18~cm measurements were made with a ground--space baseline projection at a position angle about $-35^\circ$, roughly orthogonal to the direction of the relativistic jet. Hence, if the compact feature is extended along the jet direction, it could have a lower brightness temperature. 
However, our modeling of 18~cm ground based VLBA observations (Figure~\ref{f:ground}) shows that the longitudinal extent of the most compact feature cannot exceed 0.7~mas. Accordingly, the 18~cm brightness temperature estimate might decrease by a factor of two at most.


\subsection{Visibility-based estimates of brightness temperature limits}

The limits of brightness temperature can be estimated directly from visibility amplitude measurements assuming models of the brightness distribution of the emitting region \citep{l15}. If the brightness distribution can be described by a two-dimensional Gaussian pattern, then the observed visibility amplitude $V_r$ measured at a Fourier spacing, $r_{uv}$, determines a minimum brightness temperature, $T_\mathrm{b,min}$, supported by the measurement. If both the visibility amplitude and its rms error, $\sigma_{V_r}$, are known, then a characteristic brightness temperature, $T_\mathrm{b,char}$, can be obtained assuming that the structural detail probed by the visibility measurement is resolved. This assumption translates into requiring for the zero-spacing flux density $V_0\ge V_r+\sigma_{V_r}$ \citep{l15}. Application of this approach to data from large ground-based VLBI survey programs has demonstrated that the interval ($T_\mathrm{b,min};T_\mathrm{b,char}$) brackets the brightness temperature of the most compact structure in compact extragalactic radio sources, provided that the measurements are made at Fourier spacings larger than 200 M$\lambda$. This suggests that the method is well-suited for application to \ra\ measurements made on ground-to-space baselines.

Table~\ref{t:results} lists the respective minimum and characteristic brightness temperatures of 3C\,273 in columns (11) and (12) obtained from the visibility amplitudes measured on the baselines between ground antennas and the space radio telescope of \ra. These values may differ from the estimates in column (10) for which explicit assumptions of the zero-spacing flux density were made. For every epoch the Table presents results for the most sensitive baseline.
The visibility data used for making these estimates were first coherently averaged within two-minute intervals, and then the amplitudes measured over single \ra\ scans were used to  estimate $T_\mathrm{b,min}$ and $T_\mathrm{b,char}$.
%
These limits imply that the brightness temperature of the emission detected by the \ra\ measurements cannot be smaller than about $5\times10^{12}$~K.
%

\section{Discussion}

The highest brightness temperatures implied by our measurements are at 18~cm, where we estimate $T_\mathrm{b}>4 \times 10^{13}$~K. At this wavelength, scatter broadening may be non-negligible, and estimates that account for the scatter broadening would be even higher. However, the recent discovery of refractive substructure in SgrA$^*$ \citep{gwinn_etal14} suggests that scattering in the turbulent interstellar medium can result in an overestimate of the correlated flux density on long Earth-space baselines with a corresponding overestimate of the apparent brightness temperature \citep{JG15} at 18~cm. Even if this effect is present in our data, we underline that values up to $2\times10^{13}$~K are revealed at shorter wavelengths, where scattering is weak and the effects of substructure are negligible. A detailed analysis of this effect is presented in an accompanying paper by \cite{MJ2016}, who estimate that $T_\mathrm{b} = 7 \times 10^{12}$~K at 18~cm. Thus, the inferred brightness temperatures are likely to be close to $10^{13}$~K at each of our observing wavelengths.

The apparent extreme brightness temperature in 3C\,273 reported here is difficult to explain. VLBA monitoring of 3C\,273 at 15~GHz indicates a maximum observed jet speed $\beta_\mathrm{app}\approx15$\,$c$~\citep{lister_etal13}, implying an amplification $\delta\lesssim13$, as estimated by \cite{J05,FM3}.
Considering the Compton limit of $10^{11.5}$~K or the equipartition brightness temperature value of $5\times10^{10}$~K, this amplification is at least 10 to 60 times too small to produce the observed brightness temperature at 18~cm, and it is 3 to 20 times too small to explain the 1.3~cm results.
This suggests that the radio emission from 3C\,273 may not be well described by incoherent synchrotron radiation from relativistic electrons \citep{kar00}.

We have considered the possibility that the jet velocities observed with the VLBA \citep{lister_etal13} are much lower than the bulk flow velocity relevant to Doppler boosting. However, the observed relationship between apparent brightness temperature and apparent speed of quasars follows that expected if the flow velocity and pattern velocity are related \citep{2cmPaperIV,Homan_etal06}. Another possibility is that there is sufficient injection or acceleration of relativistic electrons to balance losses due to inverse Compton cooling. However, this requires that the acceleration occurs in the region which must be several parsecs away from the central nucleus due to synchrotron self-absorption closer to the central engine \citep{BlandfordKonigl79,pushkarev_etal12}. 
Moreover, \cite{R94} has noted that sustained $\gamma$-ray luminosities would likely be required at a level that is not observed \citep[see the low level of $\gamma$-ray flux around 2013.0 in][]{ram15}. Quantitative calculations are strongly model dependent. \cite{PPM15} determined observational signatures of the Compton catastrophe and noted that they were not realized in 3C\,273.
Alternatively, some or all of the observed radio emission from 3C\,273 might be due to synchrotron radiation from relativistic protons, rather than electrons, which would raise the theoretical brightness temperature limit by a factor \citep{Jukes67} of $(m_\mathrm{p}/m_\mathrm{e})^{\nicefrac{9}{7}}\approx 1.5\times10^4$. However, this would increase the required magnetic field by a factor of $(m_\mathrm{p}/m_\mathrm{e})^2\approx4\times10^6$ that makes the proton synchrotron model problematic. Moreover, \cite{rees_proton68} has pointed out that unless the observing frequency is below the electron gyro frequency, absorption by electrons would still limit the observed brightness temperatures to the canonical values. This would imply magnetic field values of an order of $10^3$~G, far in excess of that thought to exist in quasar jets. Models considering synchrotron emission from particle-dominated electron-positron flows with a monoenergetic particle distribution \citep{TK07} can accommodate brightness temperatures of 0.5~to~$2\times10^{14}$~K, for the physical conditions measured in the jet of 3C\,273 \citep{FM3}, but due to synchrotron radiation losses, an initially monochromatic spectrum would be broadened. Coherent or collective processes such as plasma waves or stimulated synchrotron \citep{melrose99} or cyclotron \citep{BER05} emission may also lead to brightness temperatures well in excess of $10^{12}$~K.

Because our observations only detect these extreme brightness temperatures in the wide frequency range over a period of about two months~--- less than the light crossing time~--- they may also indicate the existence of a short-lived phenomenon that drives the emitting plasma far from equipartition. However, identifying such a variable region with the jet core may still be challenging due to synchrotron opacity and expected time delays between the frequencies.
Nevertheless, the core location is still more likely for this emission, as the optically thin jet of 3C\,273 has not shown very bright unresolved knots in any observations obtained from previous ground-based monitoring programs.
We note that inverse Compton loss timescale should be about one day or less at these frequencies, as estimated by \cite{slysh92,R94}.

\section{Summary}

Multi-frequency 1.7, 4.8, and 22~GHz Space VLBI observations of the \ra\ mission have revealed extremely compact structure in the jet of the archetypical quasar 3C\,273 with interferometer detections up to 13.5 Earth diameters, most probably its apparent core. The core brightness temperature is estimated to reach values in excess of $10^{13}$~K.
Sub-structure introduced by interstellar scattering \citep{MJ2016} can significantly affect estimates at 1.7~GHz. At the same time, our brightness temperature values at 4.8 or 22~GHz remain unaffected. We conclude that it is difficult to interpret the data in terms of conventional incoherent synchrotron radiation.
An ongoing analysis of a large sample of AGN at the long \ra\ baselines will help to resolve this mystery.

\acknowledgments
The \ra\ project is led by the Astro Space Center of the Lebedev Physical Institute of the Russian Academy of Sciences and the Lavochkin Association of the Russian Federal Space Agency, and is a collaboration with partner institutions in Russia and other countries. The Arecibo Observatory is operated by SRI International under a cooperative agreement with the National Science Foundation (AST-1100968), and in alliance with Ana G. Mendez-Universidad Metropolitana, and the Universities Space Research Association. The National Radio Astronomy Observatory is a facility of the National Science Foundation operated under cooperative agreement by Associated Universities, Inc. This research is partly based on observations with the 100-meter telescope of the MPIfR (Max-Planck-Institute for Radio Astronomy) at Effelsberg. This research has made use of data from the MOJAVE database that is maintained by the MOJAVE team \citep{MOJAVE_V}. We have processed and used archival VLBA observations at 18~cm (code~BG196I). We would like to extend thanks to the NRAO group responsible for developing and deploying phased-array VLBI capabilities at the newly upgraded VLA. We thank Tuomas Savolainen, Alexander Pushkarev, Richard Schilizzi, Phil Edwards, Sergio Colafrancesco, and Martin Rees for helpful discussions. 
We thank the referee for constructive comments.
This project was supported by the Russian Foundation for Basic Research grant 13-02-12103.

{\it Facilities:} \facility{\ra\ Space Radio Telescope (Spektr-R), Arecibo, GBT, VLA, Effelsberg}.


\end{document}